# Multimodal Lyrics-Rhythm Matching

Callie C. Liao
*McLean High School*
McLean, USA
cliao2025@gmail.com

Duoduo Liao
*School of Computing*
*George Mason University*
Fairfax, USA
dliao2@gmu.edu

Jesse Guessford
*School of Music*
*George Mason University*
Fairfax, USA
jguessfo@gmu.edu

*Abstract*—Despite the recent increase in research on artificial intelligence for music, prominent correlations between key components of lyrics and rhythm such as keywords, stressed syllables, and strong beats are not frequently studied. This is likely due to challenges such as audio misalignment, inaccuracies in syllabic identification, and most importantly, the need for cross-disciplinary knowledge. To address this lack of research, we propose a novel multimodal lyrics-rhythm matching approach in this paper that specifically matches key components of lyrics and music with each other without any language limitations. We use audio instead of sheet music with readily available metadata, which creates more challenges yet increases the application flexibility of our method. Furthermore, our approach creatively generates several patterns involving various multimodalities, including music strong beats, lyrical syllables, auditory changes in a singer's pronunciation, and especially lyrical keywords, which are utilized for matching key lyrical elements with key rhythmic elements. This advantageous approach not only provides a unique way to study auditory lyrics-rhythm correlations including efficient rhythm-based audio alignment algorithms, but also bridges computational linguistics with music as well as music cognition. Our experimental results reveal an 0.81 probability of matching on average, and around 30% of the songs have a probability of 0.9 or higher of keywords landing on strong beats, including 12% of the songs with a perfect landing. Also, the similarity metrics are used to evaluate the correlation between lyrics and rhythm. It shows that nearly 50% of the songs have 0.70 similarity or higher. In conclusion, our approach contributes significantly to the lyrics-rhythm relationship by computationally unveiling insightful correlations.

## I. INTRODUCTION

In recent years, research interest has increased in utilizing Artificial Intelligence (AI) technologies for music. Even though there is heightened attention, songwriters and composers' music cognition and intuition are often overlooked in AI music research. In music, there is a variety of musical elements—rhythm, melody, harmony, lyrics, dynamics, instrumentation, timbre, etc.—that are combined to create compelling music. In particular, lyrics, through the art of literature and the diction of a vocal performance, can express emotion to the listener. When examining lyrics, keywords (important words within the text) and musically stressed syllables (syllables that the composer has emphasized) help establish musical understanding. Musically stressed syllables may differ from dictionary-defined stressed syllables. Additionally, some languages such as

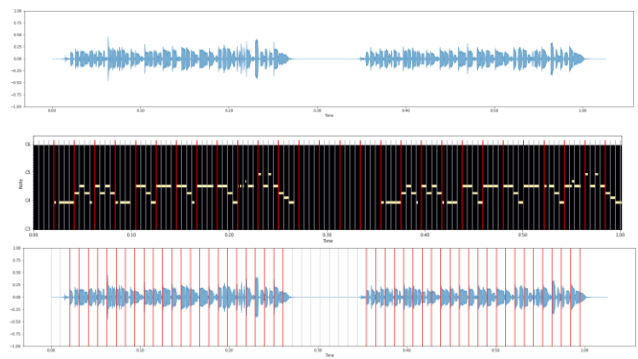

Fig. 1. The multimodal lyrics-rhythm matching. The first plot is a song's waveform. The second plot shows the first music strong beats (i.e., downbeats) in red and all other beats in gray over the piano roll where both include the rest period. The third plot indicates the matched keyword-strong-beat pairs in red over the waveform where the strong beats in gray color include the rest period.

Chinese, Japanese, and Korean lack stress [1]. Thus, in our view, keywords are preferred over stressed syllables because they encompass more forms of literary expression, which aids musical creativity and analytics.

When the lyrics are mapped onto a rhythm, the significance of keywords is reflected. Rhythm is defined as "the organization of musical events in time" [2] and creates musically metrical stress. In Fig. 2, the beats within the music measure are split into a group receiving metrical stress (strong beats) and a group not receiving metrical stress (weak beats). Strong beats usually indicate the beginning of a measure or phrase and convey the forward energy of musical emotion, although some time signatures, such as 4/4, have 2 or more strong beats within a measure. Since lyrics and rhythm work together to create musical meaning, studying the correlation between the two can serve as groundwork for further multimodal AI music research. So, we investigate the potential correlation between lyrical keywords and musical strong beats as well as between non-keywords and weak beats through audio with lyrics, which has not been studied in quantitative analytics.

There are challenges along the way, however. Misalignment between the music beats and lyrical syllables becomes unavoidable because music is often produced by humans. A singer sometimes does not perform the rhythm mechanically, which creates challenges when aligning the audio to Musical Digital Instrument Interface (MIDI[1]) files or other digital sheet music files. Moreover, some current dictionaries for splitting words into syllables are occasionally unable to produce correct syllables, and some do not produce any corresponding syllables for words unknown to the dictionaries (both new and existing). In particular, new words are a challenge since they are being invented at a fast-growing rate as technology escalates. Furthermore, lyrics are a type of free-form figurative language that contain hidden connotations and potential keywords that are difficult to determine in the text. Therefore, in general, it is challenging for current Natural Language Processing (NLP) techniques to process lyrics successfully.

---

[1]*https://www.midi.org/specifications*

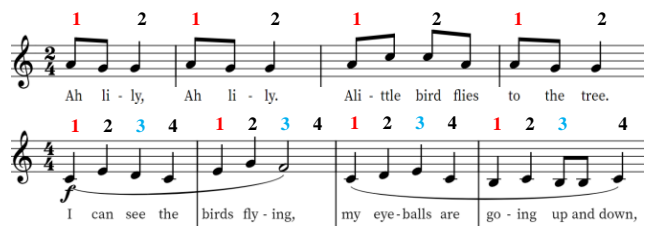

Fig. 2. Demonstrations of strong beats in music. (a) The downbeats (i.e., first strong beats) in "A Little Bird" [3] in 2/4 time signature. The downbeats are noted as 1 in red, and the weak beats are noted as 2. (b) The first and second strong beats in "Birds are Flying" [4] in 4/4 time signature. The downbeats are noted as 1 in red, and the second strong beats are noted as 3 in blue. The weak beats are noted as 2 and 4.

Above all, to conduct successful AI music research, computing technologies as well as deep knowledge and familiarity with music, literature, linguistics, and cognitive science are required. This is perhaps the most critical challenge for researchers. Therefore, making breakthroughs in any of these challenges will contribute to AI music fields.

We propose an innovative multimodal lyrics-rhythm matching approach. Our main contributions are as follows:

- Our approach computationally unveils strong, insightful correlations between lyrics and rhythm through data analytics.
- Our approach is a generalized method that determines lyrics-rhythm matching based on keywords without any language limitations. It indicates that keywords have stronger associations with strong beats. This can be employed in any language including some that lack stressed syllables.
- Creatively patterning lyrics-music multimodalities and using various metrics to measure their relationships in our approach can expand horizons in music cognition and computational analysis.
- Our approach enhances the syllabication and syllabic stress pattern identification for unknown words, which helps computational linguistics.
- Audio alignment in music continues to pose many challenges. Our research contributes to the area through the development of several novel, efficient rhythm-based audio alignment algorithms.
- Our research uses challenging audio instead of sheet music, which makes our solutions more flexible and more applicable to AI music research.
- Our research can serve as a basic building block for AI music technologies, including song identification, singing voice analysis, music structure analysis, etc.
- Our research can inspire connections between languages of music and linguistic representations in neuroscientific research, which are perhaps some of the most important interactions between language and music.

## II. RELATED WORK

Research on the computational analysis of music cognition and musicology are not addressed unlike other papers in Music Information Retrieval (MIR) [5]. As the Linguistic Stress Rule states, composers prefer to align strong beats with stressed syllables of text [1][6]. However, surprisingly few quantitative analyses can be found to support it. A few researchers conducted some basic statistical analysis and concluded that stressed syllables tend to fall on relatively strong beats of the meter in vocal music for languages with lexical stress such as English and German [7]. Another study on corpora of vocal French music has shown a strong correlation between the stress levels of syllables and their metrical strength [8]. The paper [9] presents an observational study for English popular music that performs minimal preprocessing on MusicXML[2] sheets, including monosyllabic stopwords [10] removal. It draws conclusions based on various correlations between categorical variables such as syllable stress, metric position (i.e., types of beats), and stopwords. It also helps provide fundamental statistical support to musicology and music-cognition research as well as a quantitative correlation between lyrics and melodies. However, this paper has some limitations. Firstly, it utilizes MusicXML files, so beat, lyrical syllable, rhythm, and pitch information have already been stored inside of the files; thus, extra processing is not needed, which causes additional limitations on the scope of its research. Secondly, the paper seeks for a general association between musically accented notes and stressed syllables. Thirdly, just monosyllabic stopwords removal is performed on its data to find the correlation between stopwords and non-salient notes. And fourthly, it lacks consideration for a variety of time signatures in their methods other than 4/4.

Additionally, audio alignments are among the most important steps in audio processing for MIR. Audio alignment has several distinct types, including audio-to-audio, lyrics-to-audio, audio-to-score, audio-to-visual, etc. Many researchers have conducted relevant studies in this field [11][12], including deep learning [13] based audio alignment in [14][15]. Yet, there are very few people involved in rhythm-based audio-to-audio alignment and lyrics-to-audio alignment, which are all under rhythm alignment. The paper [16] proposes a multi-task learning approach for lyrics alignment that improves the alignment accuracy through the incorporation of pitch and the integration of boundary detection in the forced-alignment algorithm. However, their performance was improved at the cost of efficiency. Another paper [17] utilizes a multi-scale neural network based on end-to-end audio-to-character architecture. It improves the alignment accuracy by predicting the character probabilities end-to-end from raw audio but only if such large datasets are available. In fact, most deep learning technologies (including the mentioned above) need large datasets for sufficient training.

---
[2]https://www.musicxml.com/

Our approach resolves the above limitations and expands the scope by using multimodal data such as music accompaniment audio, singing vocal audio, and separate lyrics text data. We perform simple, efficient, and accurate audio processing based on multimodal data, including rhythm-based audio-to-MIDI alignment and lyrics-to-audio alignment, without the use of deep learning techniques and the need for big training datasets. Most importantly, we investigate a correlation between the keywords and music strong beats due to the keywords' heightened significance when compared to other words. This correlation is applicable to other languages as well, including any language that lacks stressed syllables. Although some researchers have mentioned the lack of stressed syllables in particular languages [1], the importance of keywords in songs was not highlighted. Moreover, we use customized word syllabication and keywords extraction methods as well and consider all time signatures, including simple, compound, and asymmetric. Thus, our research places emphasis on the impact of lyrical keywords in music, making ours different from current related work.

### III. THE METHODS

The proposed approach aims to investigate positive connections between lyrics and rhythm. In this study, we focus on studying correlations among lyrical keywords and music strong beats as well as non-keywords and weak beats, and comparing the keywords approach to the stressed syllables approach to provide further insights into composers' thought processes. The architectural framework of the lyrics-rhythm matching approach is illustrated in Fig. 3. The highlighted framework is demonstrated in Fig. 1. The framework consists of three major components: music *pronunable* patterning, lyrical syllabic patterning, and lyrics-rhythm matching. Since each vocal "syllable" is recorded as a change in pronunciation, we denote each vocal "syllable" as a *pronunable* in this study.

In the framework, the first component consists of data input, beat tracking and patterning, rhythm alignment, pronunable (i.e., singing vocal "syllable") locating, and patterning with music strong beats. The second component contains syllable splitting, syllabic stress pattern identification, pronunable and lyrical syllable matching, keyword extraction, and keyword patterning. The third component targets to match keywords and strong beats and seek insights through data analytics.

The system receives lyrical texts and audio containing the singing voice and music accompaniment as input. Then, all vocal and music beats are tracked and aligned with each other. Pronunables are retrieved from the singing vocal audio and then patterned with the strong-beats based on rhythm. The keywords and syllabic stress patterns are extracted from the lyrics afterwards. The lyrical syllables need to match with the retrieved pronunables to bridge the lyrics and the beat information. Finally, the keyword strong-beat matching and overall lyrics-rhythm matching are checked using the keyword pattern and the strong-beat pattern of each pronunable associated with a word.

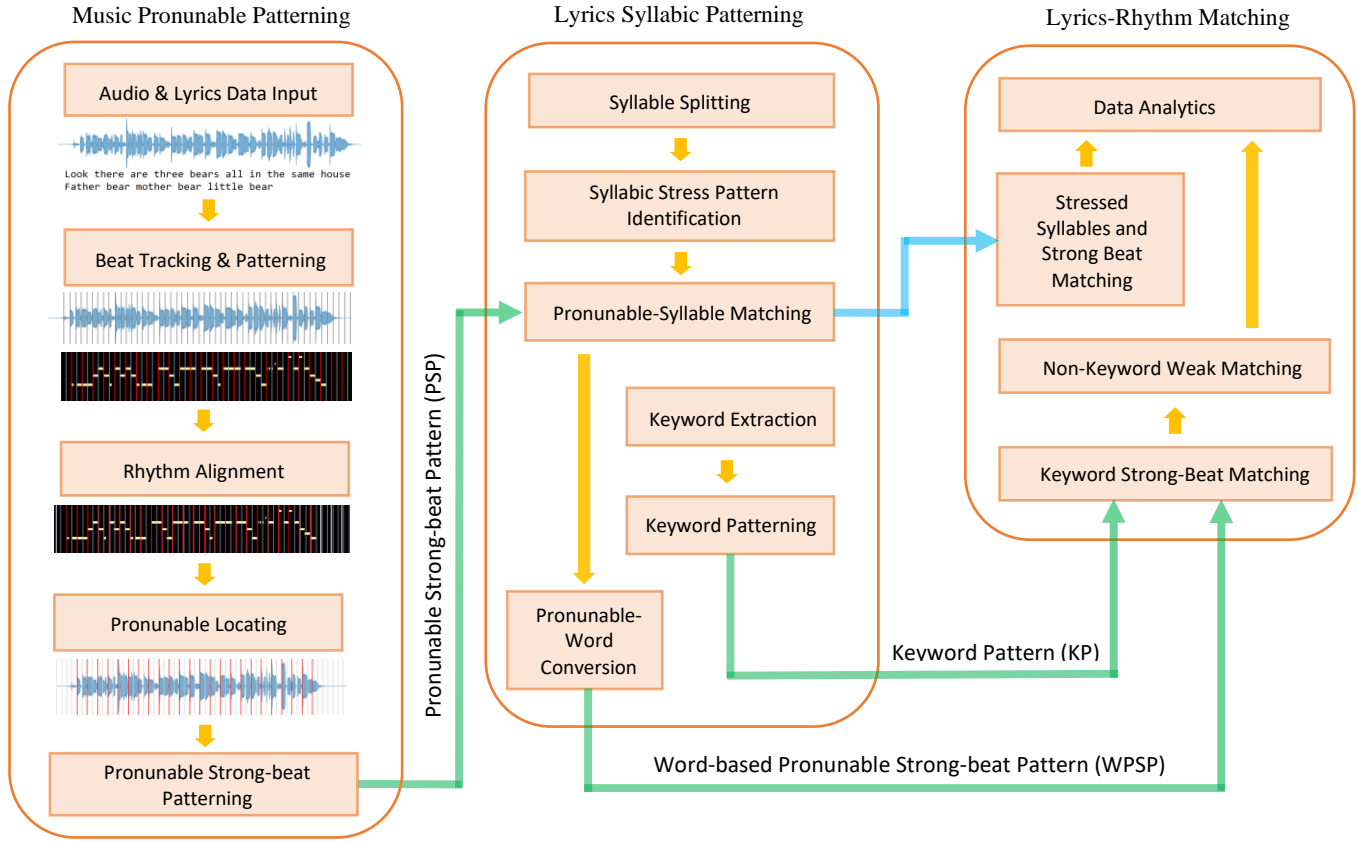

Fig. 3. The framework of Keyword Strong-Beat Matching

In the following subsections, we explain the components of the architecture in greater detail.

### A. Music Pronunable Patterning

#### 1) Beat Tracking and Strong Beat Patterning

In this framework, the input audio pair has two separate forms: a solo-singing vocal audio and an accompaniment MIDI. Singing vocal beats and music beats are tracked from these two data files, respectively. That is, two time series record onset locations (i.e., timestamps) of all tracked beats for vocal and music audios, respectively. Beat tracking is the task of identifying and synchronizing with the basic rhythmic pulse of a piece of music [18]. It has been extensively studied in MIR. Since our research focuses on lyrics-rhythm correlation findings, some existing music tools, such as Librosa[3], are used for beat tracking in this study.

After beat tracking, a Strong-beat Pattern (SP) list with its length as the total number of music beats is created to store the flags of the corresponding strong music beat locations. If a beat is a strong beat, the flag is set to 1. Otherwise, the flag is set to 0. The number of strong beats and their positions in one measure can be deduced from the time signature. Our methods can calculate the numbers for all distinct types of beats and their positions based on time signatures.

#### 2) Rhythm Alignment

In this study, rhythm alignment includes rhythm-based audio to-MIDI alignment and lyrics-to-MIDI alignment (i.e., lyrics-to-accompaniment-audio alignment).

Due to the unavoidable imperfections of a human singer, which causes delays or early starts, the vocal rhythm of the singing audio needs to be aligned with the music rhythm of the accompaniment MIDI, which is the audio-to-MIDI alignment. More specifically, the audio-to-MIDI alignment is based on two sequences, the vocal audio and music audio sequences, and uses a timing threshold or margin of error, which is set based on the tempo and time signature from the vocal audio. Additionally, the length of tracked vocal beats may be different from the length of tracked music beats. This could be due to the possibility of unmatched or missing beats. These types of beats are handled using linear interpolation. After alignments, an index list of vocal beats aligned with music beats is generated.

As for the rhythm-based lyrics-to-MIDI alignment approach, it is more complicated and consists of several steps: pronunable retrieval and locating, pronunable-beat patterning, syllable pattern identification, syllable pronunable matching, and lyrics-to-rhythm matching. These steps will be discussed further in the following sections. This approach, the rhythm-based lyrics-to-audio alignment, can be applied to any audio other than MIDI.

#### 3) Pronunable Retrieval and Locating

In the vocal audio, pronunables represent every change in pronunciation regardless of whether it is an actual syllable of a word. This includes singing words such as "woah", "nei", etc., as well as words with multiple syllables pronounced as one such as "o'er", which represents "over". Such cases present challenges for lyrical analytics, as lyrics differ from speech and other forms of literature. For this study, we only focus on retrieving the location (i.e., start time and end time) of each pronunable instead of recognizing syllables or words.

To retrieve pronunables on music beats, the pronunables associated with vocal beats are first retrieved from the vocal audio. Once all pronunables' locations are retrieved, the vocal beats need to align with the corresponding music beats in the accompaniment audio through the use of pronunables. The algorithm of pronunable locating is shown in Algorithm 1.

Since the music and voice audio do not match precisely, a time buffer or margin for error is needed for alignment. If the music beats are within the time buffer, then the beats and the beat indices are stored accordingly. Moreover, to optimize the algorithms, vocal-audio-to-MIDI alignment is processed at the same time with pronunable locating on music beats. Specifically, for each time $t_1$ in the pronunable time series $T_1$, the times within the dynamic buffer are sought in the music beat time series $T_2$ and then stored into a list $T_s$. Their corresponding indices are collected into another list $L_s$. In the time list $T_s$, for each time $t$, a certain distance function (e.g., Euclidean distance) is employed to calculate the distance between $t$ and $t_1$. Once distance calculations for all the times in $T_s$ are complete, the minimum distance $t^*$ is selected. Then, the corresponding index $i^*$ of $t^*$ is used to point to the corresponding music beat index $i_p$ for the pronunable. $i_p$ is inserted into the output index list $L$ of music beats aligned with pronunables. If no beat is found, the corresponding beat is likely unmatched or missing, so the beat is linearly interpolated.

---

**Algorithm 1**: Pronunable Locating

---

**Input**: a vocal pronunable time series $T_1$,
       a vocal beat time series $T_2$,
       a vocal-music alignment index list $L_a$,
       a time buffer $dt$.
**Output**: an index list of music beats aligned with pronunables.
1: Initialize $L$ as an empty list.
2: **for** each $t_1$ in $T_1$ **do**
3:     $(T_s, L_s) = Search(t_1, dt, T_2)$
4:     **if** $T_s$ is not empty **Then**
5:         **for** $i = 0, 1, ..., length(T_s)$ **do**
6:             $D[i] = Distance(t_1, T_s[i])$
7:         **end for**
8:         $t^* = min(D[i])$ where $i = 0, 1, ..., length(T_s)$
9:         Get corresponding index $i^*$ of $t^*$
10:       **if** $L_a$ is not empty and $i^* \in [0, length(L_a)-1]$ **Then**
11:           $i_p = L_a[i^*]$
13:       **else** // unmatched or missing
14:           $i_p = LinearInterpolation(i)$
15:       **end if**
16:       Insert $i_p$ into $L$
17:     **end if**
18: **end for**
19: **return** $L$

---

To be more mathematically specific, the target is to find the music beat that minimizes the distance from the pronunable location to match the pronunable with the closest music beat:

$$\underset{t_s \in T_s}{\arg\min} \; Distance\,(t_1, t_s) \qquad (1)$$

---

[3] https://librosa.org/

Several different distance functions can be used to calculate $D[i]$, where $i = 0, 1, ..., length(T_s)$. In this study, the Manhattan distance is selected for the implementation of Algorithm 1.

Given two points $A$ and $B$ in $n$-dimensional space such that $A = (a_1, a_2, ..., a_n)$ and $B = (b_1, b_2, ..., b_n)$, the Manhattan distance between $A$ and $B$ is defined as:

$$D(A,B) = ||A - B|| = \sum_{i=1}^{n} |a_i - b_i| \quad (2)$$

Since the beat time series is a sorted time sequence, many search algorithms such as binary search [19] can be applied in Algorithm 1. Finally, a time series representing locations of aligned pronunables is generated accordingly.

*4) Pronunable Strong-Beat Patterning*

Upon the completion of rhythm alignment and pronunable locating on music beats, the output list contains all the beat indices where the pronunables are located. Based on this index list and the SP list created through the music strong beats, a new patterned list is created to store the Pronunable Strong-beat Pattern (PSP). The PSP contains the noted locations of all the pronunables on a strong beat. If a pronunable has the same location as a strong beat, it is set to 1 in the PSP; otherwise, it is set to 0.

*B. Lyrical Syllable Patterning*

In this component, we utilize customized word syllabication and keywords extraction methods to find more lyrical keywords.

*1) Syllable Splitting and Stress Pattern Identification*

Firstly, special preprocessing for lyrics is conducted due to the potential occurrence of cases similar to pronunables. Moreover, some additional critical cases also impact the splitting of syllables. For example, despite having no syllabic contributions, the parts after the apostrophe in contractions such as "re" in "they're" and "t" in "don't" are still assigned a syllable by pronouncing dictionaries (e.g., the Carnegie Mellon University (CMU) pronouncing dictionary[4]). As a result, a customized list of words and/or letters for lyrics preprocessing is created for these types of situations. Then, utilizing a pronouncing dictionary produces one or more variants of a word's stress pattern consisting of flags. For example, in the CMU dictionary, the stress pattern is expressed through 0s and 1s. The 1s represent the stressed syllables, whereas the 0s represent the unstressed. To simplify the process in this study, only the first most common variant is appended to a list containing all the words' stress patterns. Occasionally, the dictionary may not recognize certain vocabulary and newly created words. It also may create an incorrect syllable split or stress pattern. Thus, after performing a basic stemming [20] for words such as ones that end in "ed", words are split into syllables through vowel consideration instead. As a result, stress patterns are created for such words.

The syllabic stress patterns are identified. The algorithm of syllabic stress pattern identification is shown in Algorithm 2. The output of the algorithm is a string list of stress patterns with each stress pattern being composed of 0s and/or 1s.

*2) Pronunable-Syllable Matching*

The syllable-pronunable matching is then conducted. The total number of lyrical syllables is counted by adding up each stress pattern's length. Each syllable cannot be individually compared to the pronunables, because each pronunable expresses every change in the singer's pronunciation in the most pronounceable way. As a result, the spelling of pronunables may vary drastically for each syllable. For example, the syllables of "very" are "ve" and "ry". However, "ry" can be pronounced as "ree" and "rye", which both are pronunables. In this case, the pronunable for "ry" in "very" is "ree", yet if "ree" and "ry" are compared to each other, the two will not match. Thus, comparing the total number of syllables in the lyrics to the actual number of pronunables in a song is more accurate for the matching between pronunables and lyrical syllables. This is the key step in rhythm-based lyrics-to-audio alignment as mentioned in Section A Rhythm Alignment.

---

**Algorithm 2**: Syllabic Stress Pattern Identification

**Input**: Lyrics text *S*.
**Output**: A string list of stress patterns.
1: Initialize *L* as an empty list.
2: *S'* = Preprocessed *S* with the special lyrics cases.
3: A list *T* = *Tokenize(RemovePunctuation(S'))*
4: **for** each *token* in *T* **do**
5:    Get *token*'s stress pattern from the dictionary *D*
6:    **if** stress pattern exists **Then**
7:      Append 1st stress variant as a string to *L*
8:    **else**
9:      **if** *token*'s end has no syllabic meaning **Then**
10:        Stem the *token*
11:      **end if**
12:      A list $L_s$ = *WordSyllabication(token)*
13:      Set the first syllable pattern *s* to 1
14:      Append *s* to *L*
15:    **end if**
16: **end for**
17: **return** *L*

---

*3) Keyword Extraction and Patterning*

The lyrical keywords in the text are extracted through NLP techniques. Additionally, the patterns of the keywords are identified and used to build the Keyword Pattern (KP) flag list. Since KP is now word-based yet PSP is still pronunable-based, PSP needs to be converted to a word-based pattern, i.e., Word-based Pronunable Strong-beat Pattern (WPSP). WPSP is then used to compare with KP. This is one of the key components for finding the correlation between the keywords and strong beats, not the stressed syllables.

*C. Lyrics-Rhythm Matching*

Finally, two patterns, WPSP and KP as shown in Fig. 3, are checked for whether the lyrics match the rhythm, including whether keywords land on strong beats, etc. There are multiple ways to find whether the lyrics match the rhythm, including the utilization of time series similarity and conditional probability.

*1) Conditional probability*

Clearly, the conditional probability [21] can be used for checking the matching probability of lyrics and rhythm. For example, to check if keywords land on strong beats, let *A* represent the keywords, let *B* represent the strong beats, let $P(A \cap B)$ represent the probability of overall keywords landing on strong beats, let $P(B)$ represent the probability of strong beats occurring out of all beats, and let $P(A \mid B)$ represent the likelihood of keywords occurring on the occurrence of strong beats. According to the conditional probability formula, $P(A \mid B)$ can be calculated as

$$P(A \mid B) = \frac{P(A \cap B)}{P(B)}. \quad (3)$$

---

[4]*http://www.speech.cs.cmu.edu/cgi-bin/cmudict*

*2) Cosine Similarity*

Cosine similarity [22] is one of the similarity measures for a vector space model. It measures the cosine of an angle formed by the projection of two vectors in a multi-dimensional space. To be more specific, given two vectors, *A* and *B*, representing two binary patterns, such as KP and WPSP, respectively, the cosine similarity is utilized to measure the similarity of two patterns as shown in the following formula:

$$Similarity = cos(\theta) = \frac{A \cdot B}{||A|| \, ||B||} \quad (4)$$

In Equation (4), $||A||$ is the Euclidean norm of vector $A$ = ($a_1$, $a_2$, ..., $a_n$), defined as $\sqrt{a_1^2 + a_2^2 + \cdots + a_n^2}$. Conceptually, it is the length of the vector. Similarly, $||B||$ is the Euclidean norm of vector *B*.

The cosine similarity is a real number between 0.0 and 1.0. The higher the similarity, the smaller the angle is between two vectors, *A* and *B*, in the vector space. That is, the closer the two vectors are, the better the match. Otherwise, a lower similarity represents a lower chance of being matched with each other.

Unlike conditional probability, cosine similarity is an overall matching check. It can check both whether keywords land on strong beats as well as whether non-keywords and weak beats match at the same time. In this paper, we use both matching methods for the experiments.

## IV. EXPERIMENTAL RESULTS AND ANALYSIS

This study examines lyrics-rhythm matching in using quantitative data gathered from a systematically selected dataset of songs. Our system is scalable and can be used for large datasets.

In this experiment, we use the Children's Song Dataset (CSD) [23] to pilot our study. This dataset consists of 50 English songs and 50 Korean songs. Each song is sung in two different keys by a professional female singer stored in .wav file format, and each also includes a MIDI piano file, an English lyrics text file, and text annotations in .csv file format. For English songs, there are 100 MIDI and 100 .wav files in total. The text annotations record every change in the singer's pronunciation (i.e., pronunables) using the Korean grapheme system, which differs from both the English grapheme system and the English syllabic system. This is the reason behind our introduction of pronunables. Additionally, each song contains a silence interval.

The tempo and time signature are retrieved from the metadata of vocal files beforehand since it is already noted. From the MIDI files, only the beats are utilized. We use Pretty MIDI[5] to process the MIDI data.

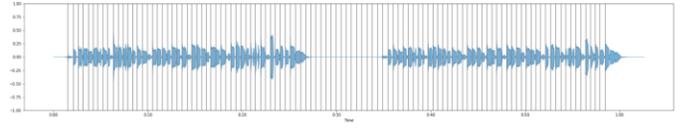

Fig. 4. The plot of beat locations over the waveform after beat tracking

Fig. 4 shows the tracked beats that are distributed over the waveform. In this example of the song file, the graph's amplitude range is (-1.0, 1.0), and the time spans over one minute. The gray lines represent every identified beat in the singing vocal audio. Based on the gray lines, some of the beats at the end are not tracked, which can be due to the duration of the last note. This does not affect our study.

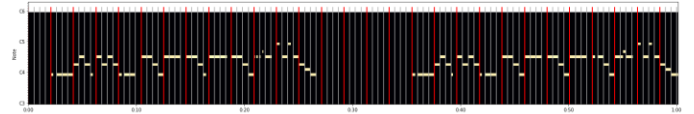

Fig. 5. Beat tracking for the music MIDI and visualization with the piano roll of notes and the song's strong beats

Likewise, beat tracking is applied to the accompaniment music MIDI file separately as shown in Fig. 5. The note range span an octave (from C4 to C5) [2]. The red lines represent the first strong beats (i.e., downbeats) from each measure and the gray lines represent all other beats. Both lines include the rest period. All the beats are perfectly aligned, which is because computer-generated MIDI files are error-free.

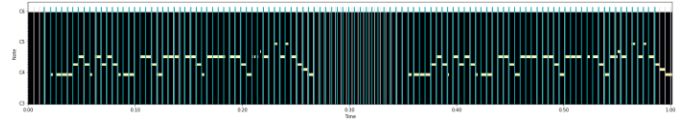

Fig. 6. The piano roll of notes and all unaligned vocal beats

In Fig. 6, both vocal beats and music beats are drawn on the same piano roll graph, which also includes the rest period. It clearly indicates that the aqua-colored beat lines appear to be misaligned with the music beats in some areas, particularly during the silent interval in this example. This shows that each slight delay in the singer's voice accumulates over time, resulting in larger misalignments until the singer is onbeat again.

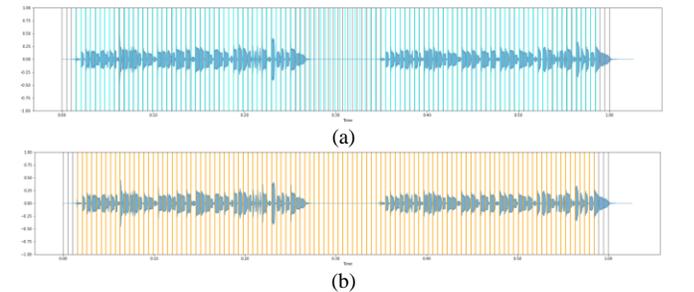

(a)

(b)

Fig. 7. Rhythm alignment. The music beats are in gray, and the unaligned vocal beats are in aqua in (a). Aligned vocal beats are in orange in (b).

In Fig. 7, (a) displays the unaligned vocal beats with music beats in the waveform and (b) shows the aligned vocal beats with music beats after rhythm alignment (i.e., rhythm-based audio-to-MIDI alignment). From the figures, our alignment algorithm has demonstrated a high accuracy.

---

[5] https://github.com/craffel/pretty-midi

In Fig. 8 (a), it indicates the original vocal beats and original pronunable locations on the singing vocal wave. The misalignment and early cutoff are still shown through the gray lines as they are from the singing vocal file. Upon completion of pronunable locating in Fig. 8 (b), the pronunable beats are matched with the music beats since every vocal beat is perfectly aligned with the music strong beats. Fig. 8 (c) further visualizes the aligned and matched pronunable beats on the piano roll with notes. The red lines indicate the first and second strong beats from the MIDI file and the gray lines indicate all the strong beats from the MIDI file, including places where no note is plotted, such as the rest period. On the other hand, the red lines only show the beats which pronunables are aligned with and therefore do not include the rest period.

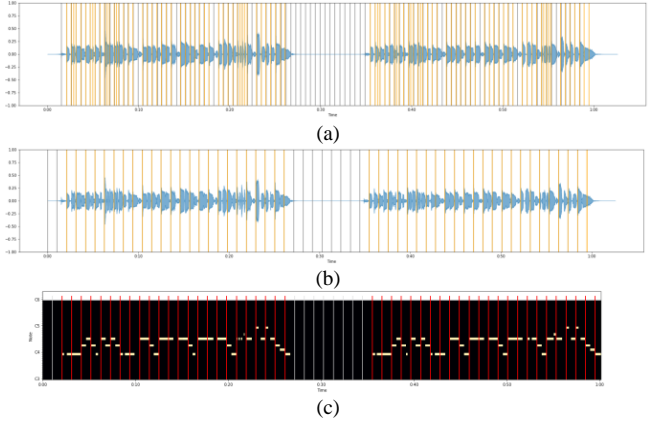

Fig. 8. Pronunable locating.
(a) The original unaligned vocal beats (gray) and pronunable beats (orange).
(b) Aligned and matched pronunable beats with music strong beats (orange).
(c) Aligned pronunable beats and all matched corresponding music strong beats (red) on the piano roll of notes.

Fig. 9 and Fig. 10 show various conditional probabilities given strong beats in several cases. Only the first key is marked, but there are 100 songs displayed in the graph. Fig. 9 describes the likelihood of the keywords occurring on the occurrence of strong beats. The bar chart compares the probabilities of each song with each other. The scatter plot presents a distribution of the probabilities over all of the song indices. The histogram displays the count of each probability range. To compare with the paper [9], we also performed a general association between musically accented notes and stressed syllables using the same dataset. Fig. 10 indicates such likelihood of the stressed syllables landing on strong beats.

More specifically, in Fig. 9 (a), based on the bar chart (a) and scatter chart (b), most of the probabilities concentrated in two clusters: Cluster 1, which is between 0.7 and 0.8, and Cluster 2, which is between 0.9 and 1.0. Additionally, Fig. 9 (c) indicates that around 12% of songs have 1.0 matching probability (i.e., all keywords landing on strong beats) and around 30% of songs have 0.9 or higher matching probabilities. Thus, it can be concluded that keywords are frequently associated with strong beats and that strong beats help emphasize keywords.

By comparing with keyword-strong-beat matching as shown in Fig. 9, Fig. 10 indicates there are two probability concentrations for stressed syllables landing on strong beats: Cluster 1, which is between 0.6 and 0.7, and Cluster 2, which is between 0.75 and 0.85. Additionally, around 20% of songs have 0.9 or higher matching probabilities. Likewise, it shows that stressed syllables are also frequently associated with strong beats and that strong beats help emphasize stressed syllables.

However, it is evident that keywords have stronger associations with Strong Beats (SB) than stressed syllables with strong beats as shown in Fig. 9 and Fig. 10 as well as summarized in Table I. This is because in many pieces of music, there are instances where the keyword lands on a strong beat, yet certain stressed syllables do not.

Our experimental results reveal an average of an 0.81 matching probability for our proposed approach, and around 30% of the songs have a match between keywords and strong beats that is at least 0.9, including 12% songs with all keywords landing on strong beats (probability = 1.0). Regarding the strong beat matching probability of 0.65 or lower, only 8% of the songs are under this probability for keywords, whereas for stressed syllables, it is 18%. We appreciate the CSD dataset, but it still has some limitations. There are occasional typos and mispronunciations in the lyrics, so our methods are sometimes unable to recognize these differences, which affects the matching accuracy.

To compare both cases for overall matching, cosine similarity in Formula 4 is utilized in this study. Fig. 11 (a) shows the overall similarity histogram for two patterns, WPSP and KP, related to keywords. Since each pattern only includes 1 or 0, it can be considered as a binary vector. 1 in the WPSP means that at least one of the pronunables of a word land on a strong beat while 0 means that all of the pronunables of a word land on at least one weak beat. But as for KP, 1 represents a keyword while 0 represents a non-keyword. In Table II and Fig. 11 (a), it clearly indicates that 46% of songs have 0.70 or higher similarity between WPSP and KP. That is, keywords tend to land on strong beats and non-keywords tend to land on weak beats for overall matching. Additionally, only around 7% of the songs have 0.55 or lower similarity between WPSP and KP, which reinforces the strong overall matching tendency.

Fig. 11 (b) shows the similarity histogram for two patterns, PSP and the Stressed Syllable Strong-beat Pattern (SSSP). Both are binary vectors. 1 in the stressed syllable pattern means a stressed syllable landing on a strong beat while 0 means unstressed syllable landing on a certain beat (a strong or weak beat). In the figure, it shows that 21% of the songs have 0.70 or higher similarity between two patterns. Furthermore, from Table II, about 28% of songs have 0.55 or lower similarity between PSP and SSSP. In comparison with Fig. 10, it can be further concluded that keywords have a stronger association with strong beats than stressed syllables.

V. CONCLUSIONS AND FUTURE WORK

This paper proposes a novel multimodal lyrics-rhythm matching approach that uses creative and efficient rhythm alignment, pronunable locating, syllabic stress identification, lyrical keyword extraction, and various patterning methods to computationally unveil strong correlations between lyrics and rhythm. Our approach can be applied to any language without limitations because we place emphasis on keywords, not stressed syllables. Moreover, our experimental results show that both stressed syllables and keywords tend to land on strong beats, with the matching between keywords and strong beats having a higher probability. But most importantly, our method uses audio instead of sheet music (e.g. MusicXML) with metadata, which is more challenging because information is not readily available. Using audio broadens our impact and extends our method to many more horizons and opportunities that are yet to be explored.

TABLE I. PROBABILITY COMPARISONS BETWEEN KEYWORDS AND STRESSED SYLLABLES LANDING ON STRONG BEATS

| Probability | Keywords w/ SBs | Stressed Syllables w/ SBs |
|---|---|---|
| Average | 0.81 | 0.77 |
| Cluster 1 | 0.70 - 0.80 | 0.60 - 0.70 |
| Cluster 2 | 0.90 - 1.0 | 0.75 - 0.85 |
| >= 0.90 | 27% of songs | 20% of songs |
| < 0.65 | 8% of songs | 18% of songs |

TABLE II. SIMILARITY COMPARISONS BETWEEN KEYWORDS AND STRESSED SYLLABLES LANDING ON STRONG BEATS

| Similarity | Keywords w/ SBs | Stressed Syllables w/ SBs |
|---|---|---|
| Average | 0.68 | 0.62 |
| <= 0.55 | 7% | 28% |
| >= 0.70 | 46% | 21% |
| >= 0.80 | 7% | 4% |

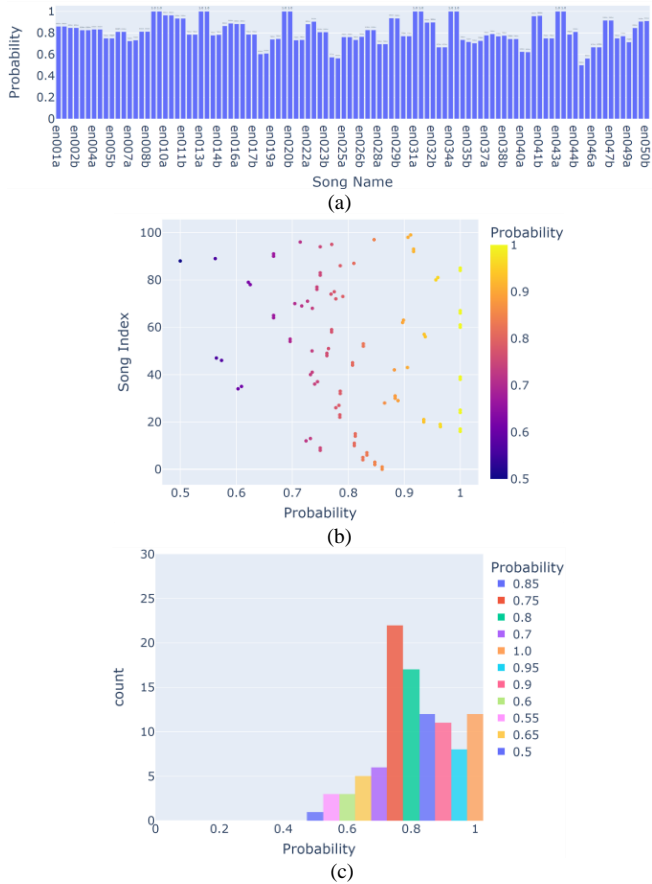

Fig. 9. The probabilities of keywords landing on strong beats (100 songs)
(a) The probabilities of keywords landing on strong beats
(b) The probability distribution of keywords landing on strong beats
(c) The probability histogram of keywords landing on strong beats

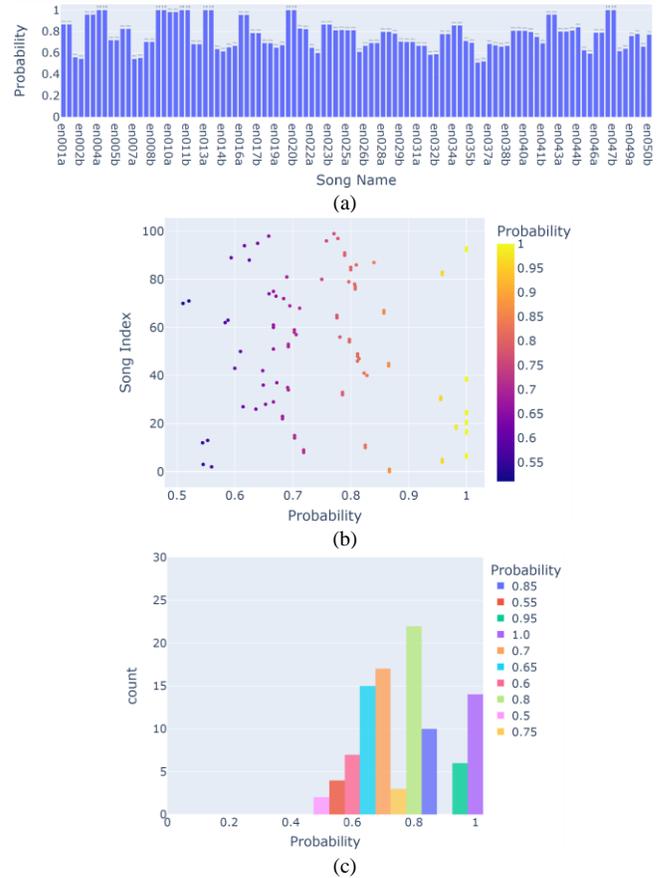

Fig. 10. The probabilities of stressed syllables landing on strong beats
(a) The probabilities of stressed syllables landing on strong beats
(b) The probability distribution of stressed syllables landing on SBs
(c) The probability histogram of stressed syllables landing on SBs

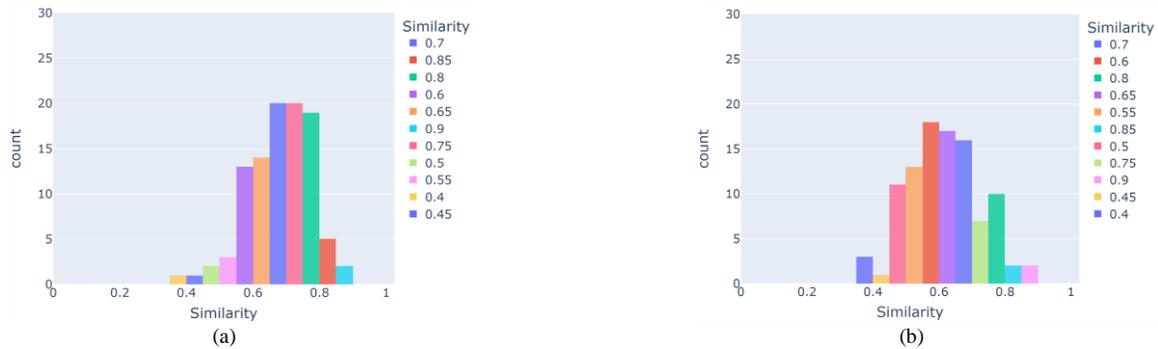

Fig. 11. The similarity histograms of KP and SSSP for all songs
(a) The similarity histogram of keywords landing on strong beats
(b) The similarity histogram of stressed syllables landing on strong beats

There are some limitations on our study, however. One limitation is our reliability on other libraries and dictionaries for some aspects of our study such as beat tracking and syllabication. In addition, there is a lack of similar audio datasets with pronunable annotations, so the size of the data used for this study is small even though our system is designed to fit the large datasets. Thus, future work includes adding singing audio recognition and automatic pronunable annotations to expand the method's use to larger and more different datasets, which reduces the overall human error as well. Furthermore, we will work on lessening and eventually completely eliminating the reliance on other libraries or dictionaries. This approach may also be used to generate more accurate music scores from singing audio because in our study, the results have shown that our establishment of a precise connection between the lyrical keywords and strong beats is accurate. Moreover, our methods can be improved upon in the future to determine the more appropriate syllabic stress pattern variant for various situations based on part-of-speech tagging using NLP techniques, as there may be multiple stress patterns for one word.